\newcommand*{\ATLASLATEXPATH}{latex/}
\documentclass[cernpreprint,texlive=2011,txfonts,UKenglish]{\ATLASLATEXPATH atlasdoc}
\pdfoutput=1
\usepackage[subfigure=true]{\ATLASLATEXPATH atlaspackage}
\usepackage[articletitle=true]{\ATLASLATEXPATH atlasbiblatex}
\usepackage{amsmath}
\usepackage{hyperref}

\usepackage{\ATLASLATEXPATH atlascontribute}

\usepackage{\ATLASLATEXPATH atlasphysics}

\graphicspath{{logos/}{figures/}}

\usepackage{placeins}
\usepackage{bm}

\renewcommand {\Bd} {\ensuremath{B^0}}
\newcommand {\Bq} {\ensuremath{B^0_q}}
\newcommand {\barBd} {\ensuremath{\bar{B}^0}}

\newcommand {\barBq} {\ensuremath{\bar{B}^0_q}}
\newcommand {\lprop} {\ensuremath{L_{\mathrm{prop}}^B}}

\AtlasTitle{Measurement of the relative width difference \\ of the $B^0$--$\bar B^0$ system with the ATLAS detector}

\author{The ATLAS Collaboration}

\AtlasRefCode{BPHY-2015-01}

\AtlasNote{ATL-COM-PHYS-2016-039}

\PreprintIdNumber{CERN-PH-2016-085}

 \AtlasJournal{JHEP}

\AtlasAbstract{%
  This paper presents the measurement of the relative width difference $\Delta \Gamma_d / \Gamma_d$
  of the $B^0$--$\bar B^0$ system using the data collected by the ATLAS experiment at the LHC in $p p$
  collisions at $\sqrt{s} = 7$ TeV and $\sqrt{s} = 8$ TeV and
  corresponding to an integrated luminosity of 25.2 fb$^{-1}$.
  The value of $\Delta \Gamma_d / \Gamma_d$ is obtained by comparing the decay-time distributions
  of $B^0 \to J/\psi K_S$ and $B^0 \to J/\psi K^{*0}(892)$ decays. The result is
$\Delta \Gamma_d / \Gamma_d  = (-0.1 \pm 1.1~\mbox{(stat.)} \pm 0.9~\mbox{(syst.)}) \times 10^{-2}$.
  Currently, this is the most precise single
  measurement of $\Delta \Gamma_d / \Gamma_d$. It agrees with the Standard Model
  prediction and the measurements by other experiments.
}

 \AtlasCoverSupportingNote{Measurement of $\Delta \Gamma_d$}{https://cds.cern.ch/record/1997304}
\AtlasCoverCommentsDeadline{3 April 2016}

\AtlasCoverAnalysisTeam{G. Borissov, E. Bouhova-Thacker, D. Scheirich, M. Skinner}

\AtlasCoverEdBoardMember{Sandro~Palestini~(CERN)~(chair)}
\AtlasCoverEdBoardMember{Adrian~John~Bevan~(London~QMUL)}
\AtlasCoverEdBoardMember{Peter~Kluit~(Nikhef)}
\AtlasCoverEdBoardMember{Camilla~Maiani~(Orsay~LAL)}

\AtlasCoverEgroupEditors{atlas-bphy-2015-01-editors@cern.ch}

\AtlasCoverEgroupEdBoard{atlas-bphy-2015-01-editorial-board@cern.ch}

\hypersetup{pdftitle={ATLAS document},pdfauthor={The ATLAS Collaboration}}

\begin{document}

\maketitle

\section{Introduction}

The width difference $\Delta \Gamma_q$, where $q=d,s$,
is one of the parameters
describing the time evolution of the $\Bq$--$\barBq$ system.
It
is defined as $\Delta \Gamma_q \equiv \Gamma_q^\mathrm{L} - \Gamma_q^\mathrm{H}$, where
$\Gamma_q^\mathrm{L}$ and $\Gamma_q^\mathrm{H}$ are the decay widths of the light and heavy
$B_q$ states, respectively.
The relative value of $\Delta \Gamma_d / \Gamma_d$
is predicted in the Standard Model (SM)~\cite{Lenz}:
\begin{eqnarray}
\Delta \Gamma_d / \Gamma_d~\mbox{(SM)}  & = & (0.42 \pm 0.08) \times 10^{-2}. \nonumber
\end{eqnarray}
Here $\Gamma_d$ is the total width of the $\Bd$ meson  defined as
$\Gamma_d = \frac{1}{2}(\Gamma_d^\mathrm{L} + \Gamma_d^\mathrm{H})$.

Measurements of $\Delta \Gamma_d$ have been performed by the BaBar \cite{dg-babar-2}, Belle \cite{DG-Belle}, and
LHCb \cite{DG-LHCB} collaborations.
The current world average value \cite{PDG-2014} is:
\begin{eqnarray}
\Delta \Gamma_d / \Gamma_d~\mbox{(World average)}  & = & (0.1 \pm 1.0) \times 10^{-2}. \nonumber
\end{eqnarray}
The current experimental uncertainty in $\Delta \Gamma_d$ is too large to perform a stringent test of the SM prediction.
In addition, independent measurements of other quantities do not constrain the value of $\Delta \Gamma_d$.
It has been shown \cite{Lenz-DG} that a relatively large variation of $\Delta \Gamma_d$
due to a possible new physics contribution would not contradict other existing SM tests.
Therefore, an experimental measurement of $\Delta \Gamma_d$ with improved precision and its comparison
to the SM prediction can provide an independent test of the underlying theory~\cite{Gershon},
complementary to other searches for new physics.

This paper presents the measurement of $\Delta \Gamma_d$ by the ATLAS experiment using
Run 1 data collected in $p p$ collisions at $\sqrt{s} = 7$ TeV in 2011 and at $\sqrt{s} = 8$ TeV in 2012.
The total integrated luminosity used in this analysis is 4.9 fb$^{-1}$ collected in 2011 and 20.3 fb$^{-1}$ collected in 2012. The value of $\Delta \Gamma_d / \Gamma_d$ is obtained by comparing the decay time distributions
of $B^0 \to J/\psi K_S$ and $B^0 \to J/\psi K^{*0}(892)$ decays. 

\section{Measurement method}
\label{sec_method}

The time evolution of the neutral $\Bq$--$\barBq$ system
is described by the Schr\"{o}dinger equation with Hamiltonian $\bm{M_q}$:
\begin{eqnarray}
\mathrm{i} \frac{\mathrm{d}}{\mathrm{d}t}
\left(
\begin{array}{c}
\Bq(t) \\
\barBq(t)
\end{array}
\right) & = & \bm{M_q}
\left(
\begin{array}{c}
\Bq(t) \\
\barBq(t)
\end{array}
\right), \nonumber \\
\bm{M_q} & = &
\left( \begin{array}{cc}
m_q & m_q^{12} \\
(m_q^{12})^* & m_q
\end{array}
\right) -
\frac{\mathrm{i}}{2}
\left( \begin{array}{cc}
\Gamma_q & \Gamma_q^{12} \\
(\Gamma_q^{12})^* & \Gamma_q
\end{array}
\right).
\end{eqnarray}
The non-diagonal elements of $\bm{M_q}$ result from the transition $\Bq \leftrightarrow \barBq$
mediated by box diagrams and depend on the parameters of the CKM quark mixing matrix.
Due to these non-diagonal elements, the $\Bq$ meson propagates as a mixture of two physical mass eigenstates
$B_q^\mathrm{L}$ and $B_q^\mathrm{H}$:
\begin{equation}
|B_q^\mathrm{L} \rangle = p |\Bq \rangle + q | \barBq \rangle,
\qquad
|B_q^\mathrm{H} \rangle = p |\Bq \rangle - q | \barBq \rangle.
\label{BLH}
\end{equation}
Here $p$ and $q$ are complex numbers satisfying $|q|^2 + |p|^2 = 1$.
$B_q^\mathrm{L}$ and $B_q^\mathrm{H}$ have distinct masses $m_q^\mathrm{L}$, $m_q^\mathrm{H}$ and widths $\Gamma_q^\mathrm{L}$, $\Gamma_q^\mathrm{H}$.
Assuming that $\Gamma_q^{12} \ll m_q^{12}$,
the following relations hold:
\begin{eqnarray}
\label{dgm}
\Delta m_q & \equiv & m_q^\mathrm{H} - m_q^\mathrm{L}  =  2 |m_q^{12}|, \\
\label{dgq}
\Delta \Gamma_q & \equiv & \Gamma_q^\mathrm{L} - \Gamma_q^\mathrm{H}  = 2 |\Gamma_q^{12}| \cos (\phi_q^{12}), \\
m_q & \equiv & \frac{1}{2} (m_q^\mathrm{L} + m_q^\mathrm{H}), \\
\Gamma_q & \equiv & \frac{1}{2} (\Gamma_q^\mathrm{L} + \Gamma_q^\mathrm{H}), \\
\phi_q^{12}  & \equiv  & \arg \left(-\frac{m_q^{12}}{\Gamma_q^{12}} \right).
\end{eqnarray}
The sign convention adopted in Eqs. (\ref{dgm}) and (\ref{dgq}) ensures that
the values of $\Delta m_q$ and $\Delta \Gamma_q$ are positive
in the Standard Model.

The decay rates of the $B_q^\mathrm{L}$ and $B_q^\mathrm{H}$ mesons
to a given final state $f$ may be different. Therefore, the
time dependence of the decay rate $\Bq \to f$ is sensitive to $f$.
The time-dependent decay rate $\Gamma (\Bq(t) \to f)$
is~\cite{Dunietz}:
\begin{equation}
\label{dr-bq}
\Gamma(\Bq(t) \to f) \propto \mathrm{e}^{-\Gamma_q t} \left[ \cosh \frac{\Delta \Gamma_q t}{2}
+ A_{\mathrm{CP}}^{\mathrm{dir}} \cos (\Delta m_q t) + A_{\Delta \Gamma} \sinh \frac{\Delta \Gamma_q t}{2}
+ A_{\mathrm{CP}}^{\mathrm{mix}} \sin (\Delta m_q t)
\right].
\end{equation}
Here $t$ is the proper decay time of the $\Bq$ meson.
The parameters $A_{\mathrm{CP}}^{\mathrm{dir}}$, $A_{\Delta \Gamma}$ and $A_{\mathrm{CP}}^{\mathrm{mix}}$
depend on the final state $f$. The abbreviations ``dir'' and ``mix'' stand for ``direct'' and ``mixing''.
By definition:
\begin{equation}
|A_{\mathrm{CP}}^{\mathrm{dir}}|^2 + |A_{\Delta \Gamma}|^2 + |A_{\mathrm{CP}}^{\mathrm{mix}}|^2 \equiv 1.
\end{equation}

Assuming that the CP-violating phase $\phi_q^{12}$ is small, which is experimentally
confirmed for both the $\Bd$ and $\Bs$ mesons \cite{phis-world},
the time-dependent decay rate $\Gamma (\barBq(t) \to f)$
is:
\begin{equation}
\label{dr-barbq}
\Gamma(\barBq(t) \to f) \propto \mathrm{e}^{-\Gamma_q t} \left[ \cosh \frac{\Delta \Gamma_q t}{2}
- A_{\mathrm{CP}}^{\mathrm{dir}} \cos (\Delta m_q t) + A_{\Delta \Gamma} \sinh \frac{\Delta \Gamma_q t}{2}
- A_{\mathrm{CP}}^{\mathrm{mix}} \sin (\Delta m_q t)
\right].
\end{equation}

The parameters $A_{\mathrm{CP}}^{\mathrm{dir}}$, $A_{\Delta \Gamma}$ and $A_{\mathrm{CP}}^{\mathrm{mix}}$
are theoretically well defined for flavour-specific final states and CP eigenstates \cite{Dunietz}.
For a flavour-specific final state $f_{\mathrm{fs}}$, such that only the decay $\Bq \to f_{\mathrm{fs}}$ is allowed
while $\bar A_f = \langle f_{\mathrm{fs}} | \barBq\rangle = 0$, the parameters are:
\begin{equation}
\label{a_flav}
A_{\mathrm{CP}}^{\mathrm{dir}} = 1,~~~ A_{\Delta \Gamma} = 0,~~~~ A_{\mathrm{CP}}^{\mathrm{mix}} = 0.
\end{equation}
For a flavour-specific final state $\bar{f}_{\mathrm{fs}}$, such that $A_f = \langle \bar{f}_{\mathrm{fs}} | \Bq\rangle = 0$,
i.e. only the decay $\barBq \to \bar{f}_{\mathrm{fs}}$ is allowed, the parameters are:
\begin{equation}
\label{a_flav1}
A_{\mathrm{CP}}^{\mathrm{dir}} = -1,~~~ A_{\Delta \Gamma} = 0,~~~~ A_{\mathrm{CP}}^{\mathrm{mix}} = 0.
\end{equation}
For the $\Bd$ decay to the CP eigenstate $\jpsi K_S$
the parameters are:
\begin{equation}
\label{a_cp}
A_{\mathrm{CP}}^{\mathrm{dir}} = 0,~~~ A_{\Delta \Gamma} = \cos (2 \beta),~~~~ A_{\mathrm{CP}}^{\mathrm{mix}} = - \sin (2 \beta).
\end{equation}
Here $\beta$ is the angle of the unitarity triangle of the CKM matrix:
\begin{equation}
\beta = \arg \left( -\frac{V_{cd}^{~} V_{cb}^*}{V_{td}^{~} V_{tb}^*} \right).
\end{equation}

If the initial flavour of the $\Bq$ meson is not tagged, the decay rates given by
Eqs. (\ref{dr-bq}) and (\ref{dr-barbq})
are added together. In this case, the production asymmetry $A_{\mathrm{P}}$
of the $\Bq$ meson in $p p$ collisions should be taken into account.
This asymmetry is defined as:
\begin{equation}
A_{\mathrm P} = \frac{\sigma(\Bq) - \sigma(\barBq)}{\sigma(\Bq) + \sigma(\barBq)},
\end{equation}
where $\sigma$ denotes the production cross-section of the corresponding particle.
Although $b$ quarks are predominantly produced in $ b \bar b$ pairs, which result in an equal
number of $b$ and $\bar b$ quarks, the presence of a valence $u$ quark in $p p$ collisions
leads to a small excess of $B^+$ mesons  (quark content $\bar b u$) over $B^-$ mesons $(b \bar u)$ \cite{Norrbin-1,Norrbin-2}. Similarly,
there is an excess of $B^0(\bar b d)$ mesons over $\bar B^0 (b \bar d)$ mesons due to the presence of a valence $d$ quark.
The larger number of $B$ mesons than $\bar B$ mesons
is compensated for by the excess of $b$ baryons
over their corresponding anti-particles.
In each case the excess is expected to be of the order of 1\%.
Only the LHCb experiment has measured $A_{\mathrm P}$ in $p p$ collisions so far \cite{ap-lhcb}.
Their result is not directly applicable to the conditions of the ATLAS experiment because of
the different ranges of pseudorapidities $\eta$ and transverse momenta $p_\mathrm{T}$ of the detected $B$ mesons.
Therefore, a dedicated measurement of $A_{\mathrm P}$ is necessary. This measurement
is presented in Section \ref{sec_asymmetry}.

Taking into account the production asymmetry $A_{\mathrm P}$ and using Eqs.~(\ref{dr-bq}) and (\ref{dr-barbq}),
the untagged time-dependent
decay rate $\Gamma[t,f]$ 
to a final state $f$ is:
\begin{eqnarray}
\Gamma[t,f] & \equiv & \sigma(\Bq) \Gamma(\Bq(t) \to f) + \sigma(\barBq) \Gamma(\barBq(t) \to f) \nonumber \\
\label{gamma_f}
 & \propto & \mathrm{e}^{-\Gamma_q t} \left[ \cosh \frac{\Delta \Gamma_q t}{2}
+ A_{\mathrm P} A_{\mathrm{CP}}^{\mathrm{dir}} \cos (\Delta m_q t) + A_{\Delta \Gamma} \sinh \frac{\Delta \Gamma_q t}{2}
+ A_{\mathrm P} A_{\mathrm{CP}}^{\mathrm{mix}} \sin (\Delta m_q t)
\right].
\end{eqnarray}

The width difference $\Delta \Gamma_q$ can be extracted from the decay time distribution of the decay
$\Bq (\barBq) \to f$ using Eqs. (\ref{a_flav})--(\ref{gamma_f}).
Such a measurement employing a single final state $f$,
although possible, would give poor precision for $\Delta \Gamma_q$.
This is because $\Delta \Gamma_q \ll \Gamma_q$ and therefore
the term $\mathrm{e}^{-\Gamma_q t}$ dominates the decay time distribution.
A more promising method
\cite{dg-babar-2}
consists in obtaining $\Delta \Gamma_q$  from the ratio of the decay time distributions of
two different decay modes of $B_q$,
one of them being a CP eigenstate
and the other a flavour-specific state. Using this ratio eliminates the dominant factor $\mathrm{e}^{-\Gamma_q t}$
and leads to improved precision for $\Delta \Gamma_q$.

The measurement of $\Delta \Gamma_d$ presented in this paper employs the ratio of
the CP eigenstate $J/\psi K_S$
and the flavour-specific states $J/\psi K^{*0}(892)$ and $J/\psi \bar K^{*0}(892)$.
The $J/\psi K^{*0}$ and $J/\psi \bar K^{*0}$ states
are added together and are denoted by $J/\psi K^{*0}$ throughout this paper,
unless otherwise specified.

The decay rate $\Gamma[t,J/\psi K_S]$ is obtained from Eqs. (\ref{a_cp})--(\ref{gamma_f}):
\begin{equation}
\label{gcp}
\Gamma[t,J/\psi K_S] \propto \mathrm{e}^{-\Gamma_d t} \left [
\cosh \frac{\Delta \Gamma_d t}{2} + \cos(2 \beta) \sinh \frac{\Delta \Gamma_d t}{2}
- A_{\mathrm P} \sin(2 \beta) \sin (\Delta m_d t) \right ].
\end{equation}
The expression for $\Gamma[t,J/\psi K^{*0}]$ is obtained from Eqs. (\ref{a_flav}), (\ref{a_flav1}), and (\ref{gamma_f})
by summing over the $J/\psi K^{*0}$ and $J/\psi \bar K^{*0}$ final states:
\begin{equation}
\label{gfs}
\Gamma[t, J/\psi K^{*0}]  \propto  \mathrm{e}^{-\Gamma_d t} \cosh \frac{\Delta \Gamma_d t}{2}.
\end{equation}
If the detection
efficiencies of $K^{*0}$ and $\bar K^{*0}$ are different,
the term proportional to $A_{\mathrm P}$ in Eq. (\ref{gamma_f}) also contributes to Eq. (\ref{gfs}).
This contribution is multiplied by the relative difference in the detection efficiencies
of $K^{*0}$ and $\bar K^{*0}$ mesons.
Both of these factors
are of the order of $10^{-2}$, which is shown in Section \ref{sec_asymmetry}.
Therefore, the contribution of the term proportional to $A_{\mathrm P}$
is of the order of $10^{-4}$
and is neglected. Another contribution to Eq. (\ref{gfs}) coming from CP violation in mixing
is experimentally constrained to be less than 0.1\% and is also neglected in this analysis.

The world average values of $\Gamma_d$, $\Delta m_d$ and $\beta$ are \cite{PDG-2014}:
\begin{eqnarray}
1/\Gamma_d & = & (1.520 \pm 0.004) \times 10^{-12}~\mbox{s}, \\
\Delta m_d & = & (0.510 \pm 0.003) \times 10^{12} ~\mbox{s}^{-1}, \\
\sin 2 \beta & = & 0.682 \pm 0.019.
\end{eqnarray}
Their uncertainties produce a negligible impact on the measured value of $\Delta \Gamma_d$.

In this analysis, the proper decay length of the $\Bd$ meson, $\lprop = c t$, 
is used in place of the proper decay time $t$.
The procedure to measure $\lprop$ is
described in Section \ref{sec_lifetime}.
The decay rates $\Gamma[\lprop, \jpsi K_S]$ and $\Gamma[\lprop, \jpsi K^{*0}]$
expressed as functions of $\lprop$ are:
\begin{eqnarray}
\label{glcp}
\Gamma[\lprop, J/\psi K_S]    & = & \int_0^\infty G(\lprop - c t, \jpsi K_S) \Gamma[t, J/\psi K_S] \mathrm{d} t, \\
\label{glfs}
\Gamma[\lprop, J/\psi K^{*0}] & = & \int_0^\infty G(\lprop - c t, \jpsi K^{*0}) \Gamma[t, J/\psi K^{*0}] \mathrm{d} t.
\end{eqnarray}
Here $G(\lprop - c t, \jpsi K_S)$ and $G(\lprop - c t, \jpsi K^{*0})$ are the $\lprop$ detector resolution functions
for the $\Bd \to \jpsi K_S$ and $\Bd \to \jpsi K^{*0}$ channels, respectively.
These functions are discussed in Section~\ref{sec_lifetime}.

The width difference $\Delta \Gamma_d$ is obtained from the fit to the ratio $R(\lprop)$ of the
distributions of the number of reconstructed $\Bd \to J/\psi K_S$ and $\Bd \to J/\psi K^{*0}$ candidates
as a function of $\lprop$.
The expected form of $R(\lprop)$ is obtained using Eqs. (\ref{glcp}) and (\ref{glfs}).
The details of the fitting procedure are given in Section~\ref{sec_rcorr}.
Many experimental systematic uncertainties
cancel in the ratio of the $\lprop$ distributions,
which improves the precision of the $\Delta \Gamma_d$ measurement.
This is an important advantage of the method used in this analysis.

A similar method is used by the LHCb Collaboration \cite{DG-LHCB}, except that
the value of $\Delta \Gamma_d$ is obtained from the difference of the partial decay widths of the $\Bd \to \jpsi K_S$
and $\Bd \to \jpsi K^{*0}$ decay modes
and the production asymmetry $A_\mathrm{P}(\Bd)$ is not taken into account.

The $J/\psi$ meson is reconstructed using the decay $J/\psi \to \mu^+ \mu^-$, which exploits
a clean selection of $J/\psi$ mesons and a highly efficient online trigger.
The trigger efficiencies in the two $\Bd$ decay channels are equal, apart from minor effects
related to differences in the decay kinematics,
as only the properties of the $\jpsi$ meson are used to trigger the events.
The $K_S$ and $K^{*0}$ mesons are reconstructed using the $K_S \to \pi^+ \pi^-$
and $K^{*0} \to K^+ \pi^-$ ($\bar K^{*0} \to K^- \pi^+$)
decay modes. The details of this reconstruction are given in Section~\ref{sec_sample}.

\section{The ATLAS detector}
\label{sec_atlas}

The ATLAS experiment \cite{atlas}
uses a general-purpose detector consisting of an inner tracker, a calorimeter and a muon spectrometer.
A brief outline of the components that are most relevant for this analysis is
given below.

The ATLAS experiment uses a right-handed coordinate system with its origin at the nominal interaction point (IP)
in the centre of the detector and the $z$-axis along the beam pipe.
The $x$-axis points from the IP to the centre of the LHC ring, and the $y$-axis points
upward. The inner detector (ID) surrounds the interaction point; it includes a silicon pixel detector
(Pixel), a silicon microstrip detector (SCT) and a transition radiation tracker (TRT). The ID is immersed
in an axial 2 T magnetic field. The ID covers the pseudorapidity
range $|\eta| < 2.5$.  The
pseudorapidity is defined in terms of the polar angle $\theta$ as $\eta = -\ln \tan(\theta/2)$.
The ID is enclosed by a calorimeter system
containing electromagnetic and hadronic sections.
The calorimeter is surrounded by a large muon spectrometer (MS)
in an air-core toroidal magnet system. The MS contains a combination of monitored
drift tubes (MDTs) and cathode strip chambers (CSCs), designed to provide precise position measurements
in the bending plane in the range $|\eta| < 2.7$. In addition, resistive plate chambers (RPCs) and thin
gap chambers (TGCs) with a coarse position resolution but a fast response time are used primarily to
trigger muons in the ranges $|\eta| < 1.05$ and $1.05 < |\eta| < 2.4$, respectively. RPCs and TGCs are also used
to provide position measurements in the non-bending plane and to improve the pattern recognition and track
reconstruction. Momentum measurements in the MS are based on track segments formed in at least two
of the three stations of the MDTs and the CSCs.

The ATLAS trigger system had three levels during Run 1: the hardware-based Level-1 trigger and the two-stage
High Level Trigger (HLT), which comprises the Level-2 trigger and the Event Filter. At Level-1, the muon trigger
searched for patterns of hits satisfying different transverse momentum thresholds using the RPCs and TGCs.
The region-of-interest around these Level-1 hit patterns then served as a seed for the HLT muon reconstruction, in
which dedicated algorithms were used to incorporate information from both the MS and the ID, achieving
a position and momentum resolution close to that provided by the offline muon reconstruction.

\section{Data sample and event selection}
\label{sec_sample}

This analysis uses the full sample of $p p$ collision data collected by the ATLAS detector in 2011 at
$\sqrt{s} = 7$~TeV and in 2012 at
$\sqrt{s} = 8$ TeV.
After applying strict data quality criteria the integrated luminosity
is 4.9~fb$^{-1}$ for the 2011 sample and
20.3 fb$^{-1}$ for the 2012 sample.

A set of dimuon trigger chains designed to select $J/\psi \to \mu^+ \mu^-$ decays is used \cite{trig1,trig2}.
It includes
numerous triggers with different muon $p_\mathrm{T}$ thresholds and additional topological and invariant mass
requirements.
Any dependence of the triggers on the proper decay time
cancels to a good approximation in the ratio $R(\lprop)$ introduced in Section \ref{sec_method}
because both the $\Bd \to J/\psi K_S$ and $\Bd \to J/\psi K^{*0}$ decays are selected with the same
set of triggers.

For a given event,
the primary vertex (PV) of the $p p$ collision producing the $\Bd$ meson is determined using good-quality tracks
reconstructed in the ID. The average transverse position of the $p p$ collisions (the beam spot)
is used in this determination as a constraint.
The beam spot is monitored continuously and is reconstructed at regular intervals
using several thousand interactions collected from many events.
The size of the beam spot for the 2012 data
is 15 $\mu$m in the plane transverse to the beam direction.
Due to the high LHC luminosity, each event containing a $\Bd$ meson is accompanied by a large number
of pile-up interactions, which occur at various $z$ positions along the beam line.
These background interactions produce several PV candidates. The selection of the primary vertex corresponding
to the $\Bd$ production point is described in Section \ref{sec_lifetime}.

The $J/\psi$ candidates are reconstructed from pairs of oppositely charged muons
with $p_\mathrm{T} > 2.5$ GeV and $|\eta| < 2.4$. Their ID tracks are fitted to a common vertex.
The $\chi^2$ of the vertex fit must satisfy $\chi^2(\jpsi)/\mbox{NDF} < 16$, where NDF
stands for the number of degrees of freedom and is equal to one in this case.
The mass of the $J/\psi$ candidate is required to be between 2.86 and 3.34 GeV.

The $K_S$ candidates are reconstructed from pairs of oppositely charged particle tracks not used in the
primary or pile-up vertex reconstruction. Each track is required to have at least one hit in either of the two silicon detectors.
The transverse momenta of the tracks must be greater than 400 MeV and have $|\eta| < 2.5$.
The pairs are fitted to a common vertex and kept if the $\chi^2(K_S)/\mbox{NDF} < 15$ (NDF = 1),
and the projection of the distance between the $J/\psi$ and $K_S$ vertices along the $K_S$ momentum
in the transverse plane is less than 44 cm.
The ratio of this projection to its uncertainty must be greater than 2.
Two additional requirements
are related to the point of closest approach of the $K_S$ trajectory to the $J/\psi$ vertex in the $xy$ plane.
The distance between this point and the position of the $\jpsi$ decay vertex
in the $xy$ plane is required to be less than 2 mm.
The difference in the $z$ coordinates of these two points must be less than 10 mm.
These requirements help to reduce the combinatorial background.
The mass of the $K_S$ candidate is required to be between 450 and 550 MeV.

The $\Bd \to J/\psi(\mu^+ \mu^-)\ K_S (\pi^+ \pi^-)$ candidates are constructed by
refitting the four tracks of the $J/\psi$ and $K_S$ candidates.
The muon tracks are constrained to intersect in a secondary vertex
and their invariant mass is constrained to the nominal $J/\psi$  mass \cite{PDG-2014}.
The two pions from the $K_S$ decay are constrained to originate from a tertiary vertex and
their invariant mass is constrained to the nominal mass of the $K_S$ meson \cite{PDG-2014}. The
combined momentum of the refitted $K_S$ decay tracks is required to point to the
dimuon vertex. The fit has $\mbox{NDF} = 6$.
The quality of the cascade vertex fit is ensured by the requirement $\chi^2(\Bd) - \chi^2(J/\psi) <  25$.
Finally, the transverse momentum of the $\Bd$ is required to exceed 10~GeV.

For the selection of $\Bd \to J/\psi K^{*0}$ candidates,
a $J/\psi$ candidate and two additional oppositely charged particles are combined together.
One particle is assigned the mass of the charged kaon and  the other
the mass of the charged pion. The transverse momentum of the kaon is required to exceed 800 MeV and the transverse
momentum of the pion must be greater than 400 MeV. Both tracks must have $|\eta| < 2.5$.
A vertex fit of the four selected tracks is performed where
the invariant mass of the two muon tracks
is constrained to the nominal $J/\psi$  mass. All four tracks are constrained to
originate from the same vertex. The fit has $\mbox{NDF} = 6$. The quality of the vertex fit is ensured by
the requirement $\chi^2(\Bd) - \chi^2(J/\psi) <  16$.
The invariant
mass of the $K \pi$ system
is required to be
between 850 and 950 MeV.
This range is slightly shifted with respect to the world average value of the $K^{*0}$ mass ($895.81 \pm 0.18$ MeV)
\cite{PDG-2014} to provide a better suppression of reflections from the $B_s \to \jpsi \phi$ decay.
The transverse momentum of the $K \pi$ pair is required to exceed 2 GeV and
the transverse momentum of the $\Bd$ candidate is required to be greater than 10 GeV.

Particle identification of charged hadrons is not used in this analysis.
Therefore, each pair of tracks is tested again
with the assignments of the kaon and pion swapped. If both assignments satisfy
the above selection criteria, the combination with the smaller deviation from the nominal $K^{*0}$ mass
is chosen. Section \ref{sec_asymmetry} gives more
details about the number of such events.
In this analysis the final states $J/\psi K^{*0}$ and $J/\psi \bar{K}^{*0}$ are not distinguished
and the definition of the $\Bd$ proper decay length discussed in Section \ref{sec_lifetime}
is not sensitive to the assignment of masses.
Therefore, the misidentification between pion and kaon has a limited impact on the result of this analysis.

None of the presented selection criteria for the $\jpsi K_S$ and $\jpsi K^{*0}$ final states
are applied relative to the primary interaction point, thus avoiding
a bias in the decay time distribution of the $\Bd$ candidates.
Different groups of particles from the same event can be included
in the $\jpsi K_S$ and $\jpsi K^{*0}$
samples. In such cases, the additional candidates contribute to the combinatorial background and
do not impact the signal yields.

For the measurement of $\Delta \Gamma_d$, the ratio
$R(\lprop)$ built from the number of the reconstructed $\Bd \to \jpsi K_S$ and $\Bd \to \jpsi K^{*0}$ decays is used.
In this ratio
the dependence of the reconstruction efficiencies of the two final states on $\lprop$
should be taken into account.
A large part of this dependence, together with the associated uncertainties, cancels in $R(\lprop)$
because the number of final particles in both decay modes
is the same and the procedure to measure
$\lprop$ described in Section \ref{sec_lifetime} is similar in the two cases.
Having similar selection criteria in the two channels
also minimises the decay-time bias.
Thus, the correction to the ratio $R(\lprop)$ is expected
to be small. Still, it cannot be eliminated completely because the hadronic tracks
in the $\Bd \to J/\psi K_S$ decay originate in a displaced $K_S \to \pi^+ \pi^-$ vertex,
whereas all four tracks in the $\Bd \to J/\psi K^{*0}$ decay
originate in a single vertex. This difference between the two channels
is the main source of the experimental bias in the ratio $R(\lprop)$,
which can be evaluated only with Monte Carlo (MC) simulation.
Using simulated events, the ratio of efficiencies
to reconstruct $\Bd \to J/\psi K_S$ and $\Bd \to J/\psi K^{*0}$ decays, $R_{\mathrm{eff}}(\lprop)$,  is obtained
as a function of $\lprop$.

The Monte Carlo samples are produced by simulating
the production and decays of $\Bd$ mesons using 
\texttt{PYTHIA 6.1} \cite{pythia6} for the 7 TeV MC samples and with \texttt{PYTHIA 8.1} \cite{pythia8} for the 8 TeV MC samples. In both cases, the underlying event, parton shower and hadronisation in 
the \texttt{PYTHIA}  simulation are tuned with ATLAS data \cite{tune-pythia}.
In all cases, the events are filtered at generator level by requiring two muons with $|\eta| < 2.5$  and
transverse momenta exceeding
2.5 GeV for the 7 TeV samples and 3.5 GeV for the 8 TeV samples.
The events are passed through a full simulation of the detector using the
ATLAS simulation \cite{atlas-sim} based on \texttt{Geant4} \cite{geant4,geant4-1}
and processed with the same reconstruction algorithms
as used for the data. All samples are produced with Monte Carlo configurations adjusted to properly
account for different conditions during the two years of data-taking.

\section{Proper decay length of the $\Bd$ meson}
\label{sec_lifetime}

The procedure adopted in this analysis to measure the proper decay length of the $\Bd$ meson
is explicitly designed to use the same input information for both the
$\Bd \to J/\psi K_S$ and $\Bd \to J/\psi K^{*0}$ channels. The aim of this approach is to reduce
the experimental bias in the ratio $R(\lprop)$.
The origin of the $\Bd$ meson coincides with the
primary vertex of the $pp$ collision.
The tracks
from the $\Bd$ candidate
are excluded in the measurement of the PV position.
The position of the $\Bd$ decay is determined by the $J/\psi$ vertex, which is obtained from the vertex fit
of the two muons.
The residual impact of the additional particles from the $\Bd$ decays is evaluated
using MC simulation and is found to be small.

The proper decay length of the $\Bd$ meson, $\lprop$, is determined
in the $xy$ plane of the detector because of the better precision compared to the measurement
in three dimensions, strengthened by the small transverse size of the beam spot.
A further advantage of measuring $\lprop$ in the $xy$ plane is the reduced dependence
on pile-up interactions.
The PV corresponding to the $\Bd$ production point is selected from all reconstructed PVs as follows.
For each PV candidate, the point of closest approach of the $\Bd$ trajectory to
the PV in the $xy$ plane is determined and the difference $\delta z$ of the $z$ coordinates of
these two points is measured.
The candidate with the minimum absolute value of $\delta z$ is selected as the $\Bd$ production vertex.
As with any other procedure of PV selection, this method is not ideal and occasionally a wrong PV
is selected due to the resolution for
the $\Bd$ momentum direction.
However, any selected PV should be
close enough to the true $\Bd$ production vertex because numerically $\delta z \sim \mathcal{O}$(1\ mm) and
both vertices are located on the beam line, which has a slope of about $10^{-3}$ in both the $xz$ and $yz$ planes.
The transverse size of the beam spot is about 15\ $\mu$m in both the $x$ and $y$ directions.
Therefore, the distance between the true vertex and the selected vertex in the $xy$ plane
is expected to be much less than the precision of the decay length measurement, which is about 100 $\mu$m. Thus,
the measurement of $\lprop$ performed in the $xy$ plane is not affected by a
wrong selection of the PV in a small fraction of events.

For each reconstructed $\Bd \rightarrow \jpsi K_{S}$ or $\Bd \rightarrow \jpsi K^{*0}$ candidate,
$\lprop$ is measured using the projection of the $\Bd$ decay length along the $\Bd$ momentum in the plane transverse
to the beam axis:
\begin{equation}
\label{lprop}
\lprop = \frac{(x^{\jpsi} - x^\mathrm{PV})p_x^{B} + (y^{\jpsi} - y^\mathrm{PV})p_y^{B}}{(p_\mathrm{T}^{B})^{2}}m_{\Bd}.
\end{equation}
Here $x^{\jpsi}, y^{\jpsi}$ are the coordinates of the $\jpsi$ vertex; $x^\mathrm{PV}, y^\mathrm{PV}$ are the coordinates of the primary vertex;
 $p_x^{B}, p_y^{B}$ are the $x$ and $y$ components of the momentum
of the $\Bd$ meson and $m_{\Bd} = 5279.61~\mbox{MeV}$ is
its mass~\cite{PDG-2014}.
The resolution of $\lprop$ is obtained from simulation and is parameterised by a double Gaussian
function. It is found to be similar for the two decay modes due to the applied procedure for the $\lprop$ measurement.
The uncertainty in this resolution is propagated into the systematic uncertainty of the $\Delta \Gamma_d$ measurement
as discussed in Section \ref{sec_syst}.
To obtain the proper decay length distribution, the range of $\lprop$ between $-0.3$ and 6 mm
is divided into ten bins defined
in Table \ref{tab2}. The selected bin size is much larger than the expected $\lprop$ resolution, which is about 34 $\mu$m.
In each bin of $\lprop$,
the number of $B^{0} \rightarrow \jpsi K_{S}$ and $B^{0} \rightarrow \jpsi K^{*0}$ decays are extracted from a binned log-likelihood fit to the corresponding mass distributions.

\begin{table}[htb]
  \begin{center}
    \begin{tabular}{c c c c c c c c c c c}
      \hline \hline
      Bin number & 1 & 2 & 3 & 4 & 5 & 6 & 7 & 8 & 9 & 10\\
      Lower edge [mm] & $-0.3$ & 0.0 & 0.3 & 0.6 & 0.9 & 1.2 & 1.5 & 1.8 & 2.1 & 3.0\\
      Upper edge [mm] &  0.0 & 0.3 & 0.6 & 0.9 & 1.2 & 1.5 & 1.8 & 2.1 & 3.0 & 6.0\\
      \hline \hline
    \end{tabular}
    \caption{Definition of the $\lprop$ bins.}
    \label{tab2}
  \end{center}
\end{table}
\FloatBarrier
In this fit,
the mass distributions are modelled by a sum of functions describing the signal and background components.
For the $B^{0} \rightarrow \jpsi K^{*0}$ channel, the signal function $f_{s}^{\jpsi K^{*0}}$ is defined as the sum
of two Gaussian functions.
The Gaussian functions are constrained to have the same mean.
The background function $f_{b}^{\jpsi K^{*0}}$ is defined using an exponential function with a
second-order polynomial as the exponent.
The fit is first applied to the total sample to determine the mean and standard deviations of the two Gaussian functions
and their relative fractions.
For the fit in each $\lprop$ bin, all parameters describing the signal, except
the normalisation of $f_{s}^{\jpsi K^{*0}}$, are fixed to the values obtained in the fit of the total sample.
It was verified in this analysis that fixing the parameters of the signal does not produce any bias in the result.
The parameters of $f_{b}^{\jpsi K^{*0}}$ remain free.

The signal function for the $B^{0} \rightarrow \jpsi K_{S}$ channel $f_{s}^{\jpsi K_{S}}$ is defined as the sum of two
Gaussian functions. The background is modelled by the sum of two functions:
$f_{b}^{\jpsi K_{S}} = f_{b}^{c} + f_{b}^{B_{s}}$.
The combinatorial background function $f_{b}^{c}$ is defined using an exponential function
with a second-order polynomial as the exponent.
The second function, $f_{b}^{B_{s}}$, accounts for the contribution from $\Bs \rightarrow \jpsi K_{S}$ decays
and is defined as the sum of two Gaussian functions. The
$\Bs \to \jpsi K_{S}$ contribution is visible in the mass distribution
as a shoulder in the signal peak. Its fraction relative to the $\Bd \to \jpsi K_S$ signal is $\sim 1\%$.
The signal Gaussian functions are constrained to have the same mean.
The relative fractions and standard deviations of the $\Bs$ background Gaussian functions are parameterised to be the same as
those of the signal Gaussian functions.
The $\Bs$ background Gaussian functions are also constrained to have the same mean.
The mean of the $\Bs$ background Gaussian functions is shifted relative to the mean of
the signal Gaussian functions by the difference between the nominal masses of the $\Bs$ and $\Bd$ mesons (87.34 MeV)
\cite{PDG-2014}.
The fit is first applied to the total sample to determine the mean and standard deviations of the signal
Gaussian functions and their relative fractions.
For the fit in each $\lprop$ bin, all parameters describing the signal, except the normalisation of
$f_{s}^{\jpsi K_{S}}$, are fixed to the values obtained in the fit of the total sample. It was verified
in this analysis that fixing the parameters of the signal does not
 produce any bias in the result. The parameters of $f_{b}^{B_{s}}$ are also fixed, except for the normalisation.
All parameters of $f_{b}^{c}$ remain free.

The separation of the $\Bd \to \jpsi K_{S}$ and $\Bs \to \jpsi K_{S}$ contributions
is important for the $\Delta \Gamma_d$ measurement because the
mean lifetimes of the $\Bd$ and $\Bs$ mesons decaying to this CP eigenstate are different. On the contrary, the separation
of $\Bd \to \jpsi K^{*0}$ and $\Bs \to \jpsi K^{*0}$ decays is not necessary
because the lifetimes of the $\Bd$ and $\Bs$ mesons decaying to this final state
are equal to within 1\% \cite{PDG-2014,phis-world}.
Thus, the small ($\sim$1\%) contribution of
the $\Bs \to \jpsi K^{*0}$ decay
does not have an impact on the $\Delta \Gamma_d$ measurement.

The fit ranges of the $\jpsi K_S$ and $\jpsi K^{*0}$ mass distributions are selected such that
the background under the $\Bd$ signal is smooth.
The mass distribution $m(\jpsi K_S)$ contains a contribution from partially reconstructed
$B \to \jpsi K_S \pi$ decays. This contribution has a threshold at $m(\jpsi K_S) \simeq 5130$ MeV. For this
reason, the fit range $5160 < m(\jpsi K_S) < 5600$ MeV is selected. The corresponding contribution of
$B \to \jpsi K^{*0} \pi$ decays is smaller. Therefore, the lower limit of the fit range of $m(\jpsi K^{*0})$
is selected at 5000 MeV. The impact of the selection of the fit range on the value of $\Delta \Gamma_d$
is included in the systematic uncertainty.

\begin{figure}[htb]
  \begin{center}
    \subfigure[]{
      \hspace*{-0.5cm}
      \includegraphics[width=0.52\textwidth]{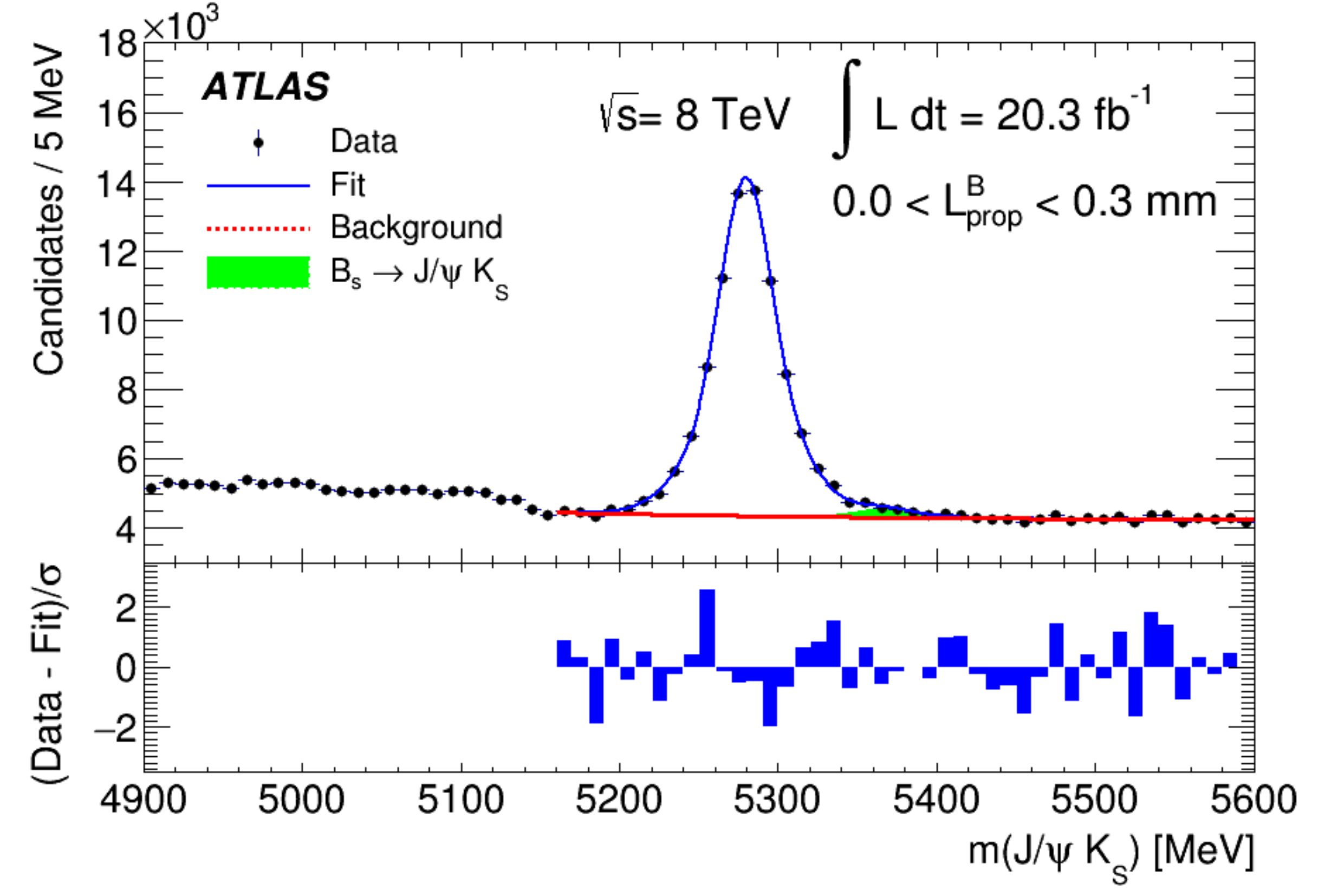}
    }
    \subfigure[]{
      \hspace*{-0.5cm}
      \includegraphics[width=0.52\textwidth]{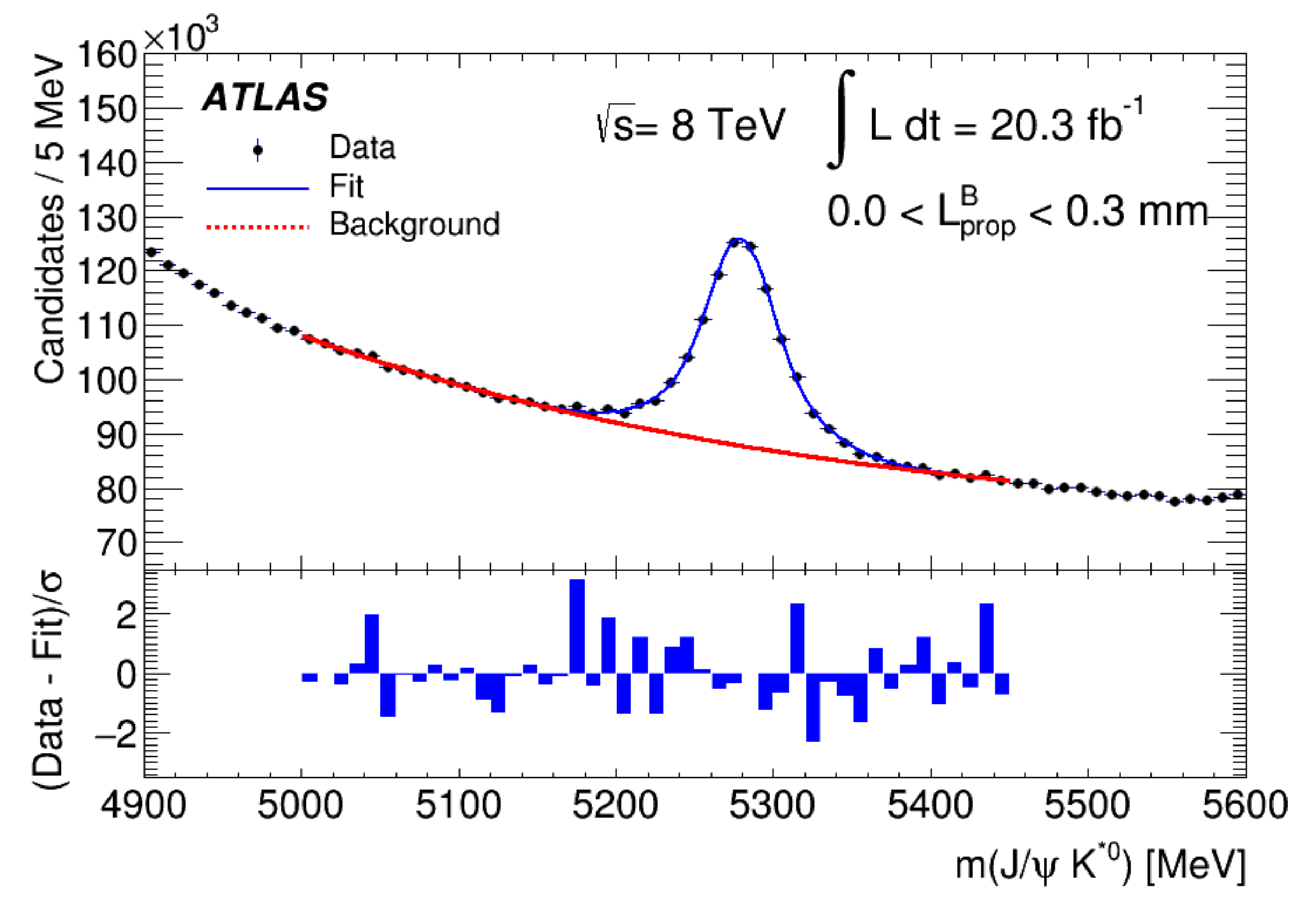}
      \hspace*{-0.6cm}
    }
    \caption{The invariant mass distributions for (a) $B^{0} \rightarrow \jpsi K_{S}$ candidates
    and (b) $B^{0} \rightarrow \jpsi K^{*0}$ candidates for the 2012 data sample for $0.0 < \lprop < 0.3$ mm.
    The full line shows the result of the fit to the function described in the text. The dashed line
    shows the combinatorial background contribution. The filled area in figure (a) shows the peaking background
    contribution from the $\Bs \to \jpsi K_S$ decay. 
    The lower frame of each figure shows the difference between each data point and the fit at that point
    divided by the statistical uncertainty of the data point.}
    \label{fig1}
  \end{center}
\end{figure}

The total number of signal $\Bd \to \jpsi K_S$ decays obtained from the fit is $28\,170\,\pm\,250$ in the 2011 data set
and $110\,830\,\pm\,520$ in the 2012 data set.
For $\Bd \to \jpsi K^{*0}$ decays the corresponding numbers are $129\,200\,\pm\,900$ in the 2011 data set
and $555\,800\,\pm\,1\,900$ in the 2012 data set.
Figure \ref{fig1} shows the fitted mass distribution of $B^{0} \rightarrow \jpsi K_{S}$ candidates and
$B^{0} \rightarrow \jpsi K^{*0}$ candidates for $0.0 < \lprop < 0.3$ mm.

The ratio of the numbers of $\Bd$ candidates in the two channels computed in each $\lprop$ bin $i$ gives
the experimental ratio $R_{i, {\mathrm{uncor}}}$ defined as:
\begin{equation}
\label{rlp}
R_{i, {\mathrm{uncor}}} =  \frac{N_i(J/\psi K_S) }
                  {N_i(J/\psi K^{*0}) }.
\end{equation}
Here $N_i(J/\psi K_S)$ and $N_i(J/\psi K^{*0})$ are the numbers of events in a given bin $i$.
This ratio has to be
corrected by the ratio of the reconstruction efficiencies in the two channels as discussed in Section \ref{sec_mc_rd}.

\FloatBarrier

\section{Production asymmetry of the $\Bd$ meson}
\label{sec_asymmetry}

The production asymmetry $A_{\mathrm P}$ of the $\Bd$ meson can be obtained from
the time-dependent charge asymmetry of the flavour-specific $\Bd \to \jpsi K^{*0}$ decay.
If the initial flavour of the $\Bd$ meson is not determined, it follows from Eqs. (\ref{a_flav}), (\ref{a_flav1}), and (\ref{gamma_f}) that the time-dependent rate of the decay $\Bd \to \jpsi K^{*0}$ is equal to:
\begin{equation}
\label{gks1}
\Gamma[t,\jpsi K^{*0}] \propto \mathrm{e}^{- \Gamma_d t } \left[
\cosh  \frac{\Delta \Gamma_d t}{2}
+ A_{\mathrm P} \cos (\Delta m_d t) \right ],
\end{equation}
while the time-dependent rate of the decay $\barBd \to \jpsi \bar{K}^{*0}$ is equal to:
\begin{equation}
\label{gks2}
\Gamma[t,\jpsi \bar K^{*0}] \propto \mathrm{e}^{- \Gamma_d t } \left[
\cosh \frac{\Delta \Gamma_d t}{2}
- A_{\mathrm P} \cos (\Delta m_d t) \right ],
\end{equation}
$CP$ violation in mixing is predicted to be small in the SM and is omitted from these expressions.

The terms proportional to $A_{\mathrm P}$ in Eqs. (\ref{gks1}) and (\ref{gks2}) reflect the oscillating component
of the $\Bd \to J/\psi K^{*0}$ decay. The corresponding charge asymmetry due to $\Bd$ oscillations
in bin $i$ of $\lprop$, $ A_{i, {\mathrm{osc}}}$, is defined as:
\begin{equation}
A_{i, {\mathrm{osc}}} \equiv \frac{
 \int_{L^{\mathrm{min}}_i}^{L^{\mathrm{max}}_i} \left ( \int_0^\infty G(\lprop - c t, \jpsi K^{*0}) (\Gamma[t,\jpsi K^{*0}] - \Gamma[t,\jpsi \bar{K}^{*0}]) \mathrm{d} t \right ) \mathrm{d} \lprop}
{\int_{L^{\mathrm{min}}_i}^{L^{\mathrm{max}}_i} \left ( \int_0^\infty G(\lprop - c t, \jpsi K^{*0}) (\Gamma[t,\jpsi K^{*0}] + \Gamma[t,\jpsi \bar{K}^{*0}]) \mathrm{d} t \right ) \mathrm{d} \lprop}.
\end{equation}
Here $G(\lprop - c t, \jpsi K^{*0})$ is the detector resolution of $\lprop$
for the $\Bd \to \jpsi K^{*0}$ channel.
The values of the lower and upper edges of bin $i$, $L^{\mathrm{min}}_i$ and $L^{\mathrm{max}}_i$, are
given in Table~\ref{tab2}. Using Eqs. (\ref{gks1}) and (\ref{gks2}), $A_{i, {\mathrm {osc}}}$
can be presented as:
\begin{equation}
A_{i, {\mathrm{osc}}}  =  A_{\mathrm P} \frac{
\int_{L^{\mathrm{min}}_i}^{L^{\mathrm{max}}_i} \left (  \int_0^\infty G(\lprop - c t, \jpsi K^{*0}) \mathrm{e}^{- \Gamma_d t} \cos (\Delta m_d t) \mathrm{d} t \right ) \mathrm{d} \lprop} {
\int_{L^{\mathrm{min}}_i}^{L^{\mathrm{max}}_i} \left ( \int_0^\infty G(\lprop - c t, \jpsi K^{*0}) \mathrm{e}^{- \Gamma_d t} \cosh \frac{\Delta \Gamma_d t}{2} \mathrm{d} t  \right ) \mathrm{d} \lprop}.
\end{equation}

In addition to $\Bd$ oscillations, the asymmetry in the number of $\jpsi K^{*0}$
and $\jpsi \bar{K}^{*0}$ events is also caused by
a detector-related asymmetry $A_{\mathrm{det}}$ due to differences in the reconstruction of positive and
negative particles.
The main source of $A_{\mathrm{det}}$  is the difference in the interaction cross-section of charged kaons with the detector material, which for momenta below 10 GeV is significantly larger for negative kaons \cite{PDG-2014}.
Therefore, the observed number of $K^{*0} \to K^+ \pi^-$ decays is larger than
that of $\bar{K}^{*0} \to K^- \pi^+$,
resulting in a positive value of the detector asymmetry $A_{\mathrm{det}}$.
This effect is independent of the $\Bd$ decay time.

The values of $A_{i, {\mathrm{osc}}}$ and $A_{\mathrm{det}}$ are diluted by misidentification
of the kaon and pion in the $\Bd \to \jpsi K^{*0}$ decay. The observed number of $\jpsi \bar K^{*0}$
events, $N(\jpsi \bar{K}^{*0})$, includes genuine
$\barBd \to \jpsi \bar{K}^{*0}$ and some $\Bd \to \jpsi K^{*0}$ decays.
The latter decay contributes because of a wrong assignment of the kaon and pion masses
to the two reconstructed charged particles, so that the decay $K^{*0} \to K^+ \pi^-$
is identified as a $\bar K^{*0} \to K^- \pi^+$.
The mistag fraction $W$ quantifies this wrong contribution to the $\jpsi \bar{K}^{*0}$ sample.
It is defined as
the fraction of true $\Bd \to \jpsi K^{*0}$ decays in $N(\jpsi \bar{K}^{*0})$.
The mistag fraction does not depend on the $\Bd$ decay time and is determined in simulation.
The obtained value is:
\begin{equation}
W = 0.12 \pm 0.02.
\end{equation}
The uncertainty of $W$ is systematic.
It takes into account possible variations of the MC simulation which describes $\Bd$
production and decay.
The simulation confirms that the mistag fraction is the same for $\Bd \to \jpsi K^{*0}$ and $\barBd \to \jpsi \bar{K}^{*0}$
decays within the statistical uncertainty of 0.4\% determined by the number of MC events. The systematic uncertainty
of the difference of the mistag fraction of the $\Bd \to \jpsi K^{*0}$ and $\barBd \to \jpsi \bar{K}^{*0}$ decays
cancels to large extent. 
Therefore, the same value of $W$ applies to candidates classified as $\jpsi K^{*0}$.

Using the above information, the expected charge asymmetry in bin $i$ of $\lprop$, $A_{i,{\mathrm{exp}}}$,
can be expressed as:
\begin{equation}
\label{asym}
A_{i, {\mathrm{exp}}} = \left( A_{\mathrm{det}} + A_{i, {\mathrm{osc}}} \right) (1-2 W).
\end{equation}
Here the factor $1-2 W$ takes into account the contribution of wrongly identified $\Bd$ decays, which
is the same for both $A_{\mathrm{det}}$ and $A_{i,\mathrm{osc}}$. The second-order
terms proportional to $A_{\mathrm{det}} A_{\mathrm P}$ are of the order of 10$^{-4}$ and
are neglected in this expression.

The observed charge asymmetry, $A_{i,{\mathrm{obs}}}$, is defined as:
\begin{equation}
A_{i, {\mathrm{obs}}} \equiv \frac{N_i(\jpsi K^{*0}) - N_i(\jpsi \bar{K}^{*0})}
{N_i(\jpsi K^{*0}) + N_i(\jpsi \bar{K}^{*0})}.
\end{equation}

\begin{figure}[htb]
  \centering
  \includegraphics[width=14cm]{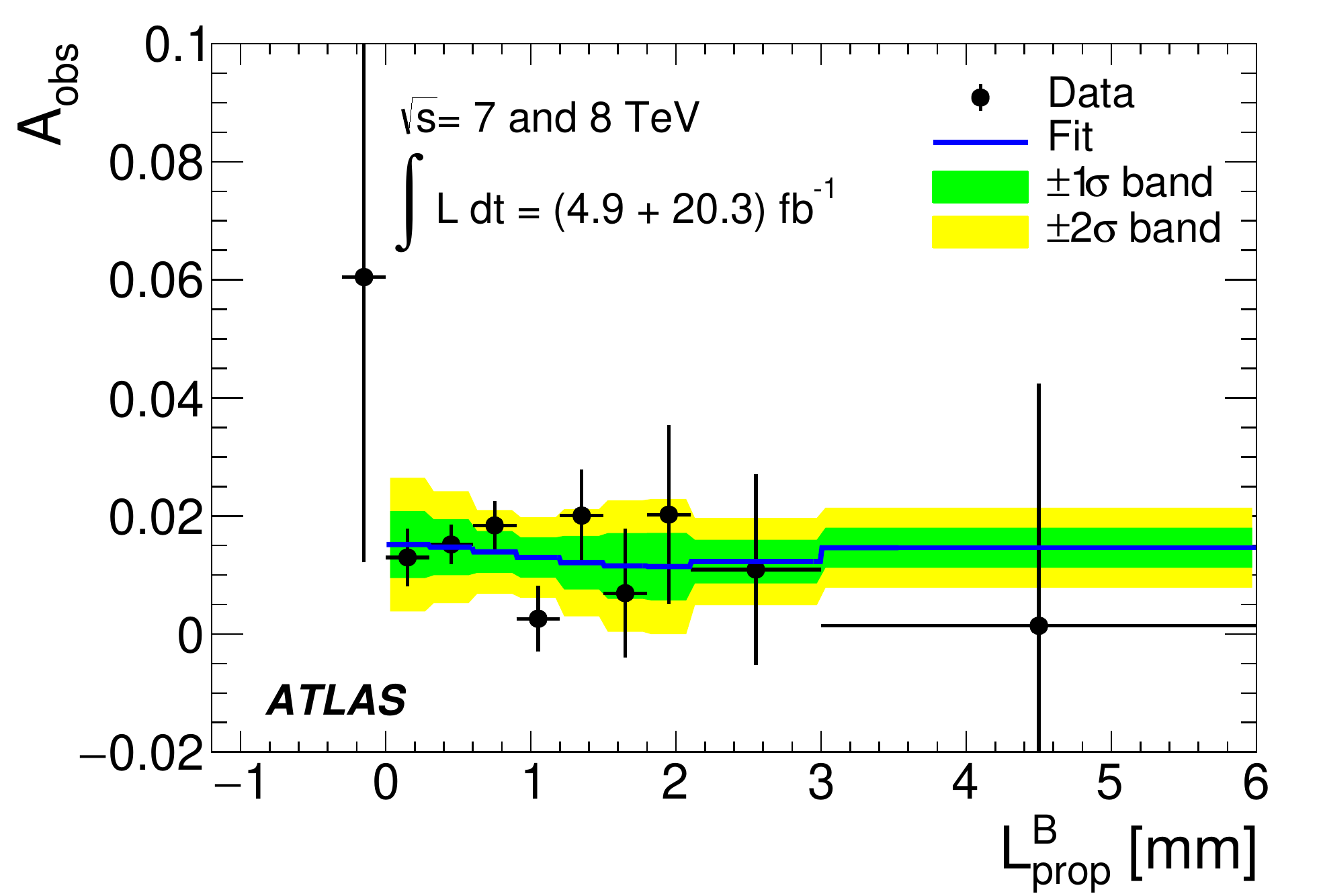}
  \caption{Observed charge asymmetry $A_{\mathrm{obs}}$ in $\Bd \to \jpsi K^{*0}$ decays
  measured as a function of the proper decay length of the $\Bd$ meson ($lprop$).
  The line shows the asymmetry $A_{\mathrm{exp}}$
  obtained from fitting Eq.~(\ref{asym}) to the data. The first point corresponding to negative proper decay length
  is not used in the fit. The error bands correspond to the combination of uncertainties obtained by the fit
  for the production asymmetry $A_{\mathrm P}$ and the detector asymmetry $A_{\mathrm{det}}$.}
  \label{fig_asym}
\end{figure}

Figure \ref{fig_asym} shows the asymmetry $A_{\mathrm{obs}}$ as a function of $\lprop$ for the 2011 and 2012
samples combined together. The result of the fit
to Eq. (\ref{asym}) is superimposed.
The asymmetry $A_{\mathrm P}$ is obtained from a $\chi^2$ minimisation:
\begin{equation}
\label{chi2_as}
\chi^2[A_{\mathrm{det}},A_{\mathrm P}] = \sum_{i=2}^{10} \frac{(A_{i, {\mathrm{obs}}} - A_{i, {\mathrm{exp}}})^2}{\sigma_i^2}.
\end{equation}
The free parameters in the fit are $A_{\mathrm{det}}$ and $A_{\mathrm P}$.
The values $\sigma_i$ are the statistical uncertainties of $A_{i,{\mathrm{obs}}}$.
The fit has a $\chi^{2}$ of 6.50 per seven degrees of freedom.
The first bin of $\lprop$ corresponds to a negative decay length due to the detector resolution.
It is not included in this sum as it is affected more than the other data points by systematic uncertainties.
Ignoring it has a negligible impact on the uncertainty of this measurement.
The fit yields the following values for the asymmetries:
\begin{eqnarray}
A_{\mathrm{det}} & = & (+1.33 \pm 0.24 \pm 0.30) \times 10^{-2}. \\
\label{apv}
A_{\mathrm P} & = & (+0.25 \pm 0.48 \pm 0.05) \times 10^{-2}.
\end{eqnarray}
In these values the first uncertainty of $A_{\mathrm P}$ and $A_{\mathrm{det}}$
is statistical and the second is due to the uncertainties in the mistag fraction
and in the deviations of $|q/p|$ from unity \cite{PDG-2014} (see Eq.~(\ref{BLH})).
The systematic uncertainty of $A_{\mathrm{det}}$ also contains a contribution from the possible
difference between the mistag fractions of the $\Bd \to \jpsi K^{*0}$ and $\barBd \to \jpsi \bar{K}^{*0}$
decays.
The value of $A_{\mathrm{det}}$ is consistent with results from simulation of interactions
in the detector.
This measurement of the $\Bd$ production asymmetry $A_{\mathrm P}$ for $p_\mathrm{T}(\Bd) > 10$ GeV and $|\eta(\Bd)| < 2.5$
is consistent with zero. It is also consistent with
the LHCb result $A_{\mathrm P} = (-0.36 \pm 0.76 \pm 0.28) \times 10^{-2}$
\cite{ap-lhcb} obtained for $4 < p_\mathrm{T}(\Bd) < 30$~GeV and $2.5 < \eta(\Bd) < 4.0$.
The measured value of $A_{\mathrm P}$ given in Eq. (\ref{apv})
is used for the extraction of the width difference $\Delta \Gamma_d$.

\section{Ratio of efficiencies}
\label{sec_mc_rd}

The ratio $R_{i, {\mathrm{uncor}}}$ given by Eq. (\ref{rlp}) is corrected by the
ratio of efficiencies $R_{i,{\mathrm{eff}}}$ computed in each $\lprop$ bin $i$. It is defined as
\begin{equation}
\label{r_eff}
R_{i,{\mathrm{eff}}} \equiv \frac{\varepsilon_i(\Bd \to J/\psi K_S)}{\varepsilon_i(\Bd \to J/\psi K^{*0})}.
\end{equation}
Here
$\varepsilon_i(\Bd \to J/\psi K_S)$  and $\varepsilon_i(\Bd \to J/\psi K^{*0})$ are the efficiencies
to reconstruct $\Bd \to J/\psi K_S$ and $\Bd \to J/\psi K^{*0}$ decays, respectively, in $\lprop$ bin $i$.
This ratio is determined using MC simulation.
To obtain reliable values for this efficiency ratio, the kinematic properties
of the simulated $\Bd$ meson and the accompanying particles
must be consistent with those in data. The comparison
of several such properties, which can produce a sizeable impact on $R_{i,{\mathrm{eff}}}$,
reveal some differences between data and simulation. Those differences are corrected for by an appropriate
re-weighting of the simulated events.

The properties taken into account include the transverse momentum and pseudorapidity of the $\Bd$ meson
and the average number of pile-up events.
The ratio of the distributions of each specified variable
in data and in simulation defines the corresponding weight.
The resulting weight applied to the MC events is defined as the product of these three weights.

The normalisation of $R_{i,{\mathrm{eff}}}$ after the re-weighting procedure is arbitrary since
only the deviation of $R_{i,{\mathrm{eff}}}$ from their average value can impact the measurement of $\Delta \Gamma_d$.
This deviation is found to not exceed 5\% for proper
decay lengths up to 2 mm. Such a stability of $R_{i,\mathrm{eff}}$
is a consequence of the chosen measurement procedure.
This stability helps to reduce the systematic uncertainty of $\Delta \Gamma_d$ due to the uncertainty of the $R_{i,{\mathrm{eff}}}$
value.

\FloatBarrier

\section{Fit of $\Delta \Gamma_d$}
\label{sec_rcorr}

\begin{figure}[htb]
  \begin{center}
    \subfigure[]{
      \hspace*{-0.5cm}
      \includegraphics[width=0.52\textwidth]{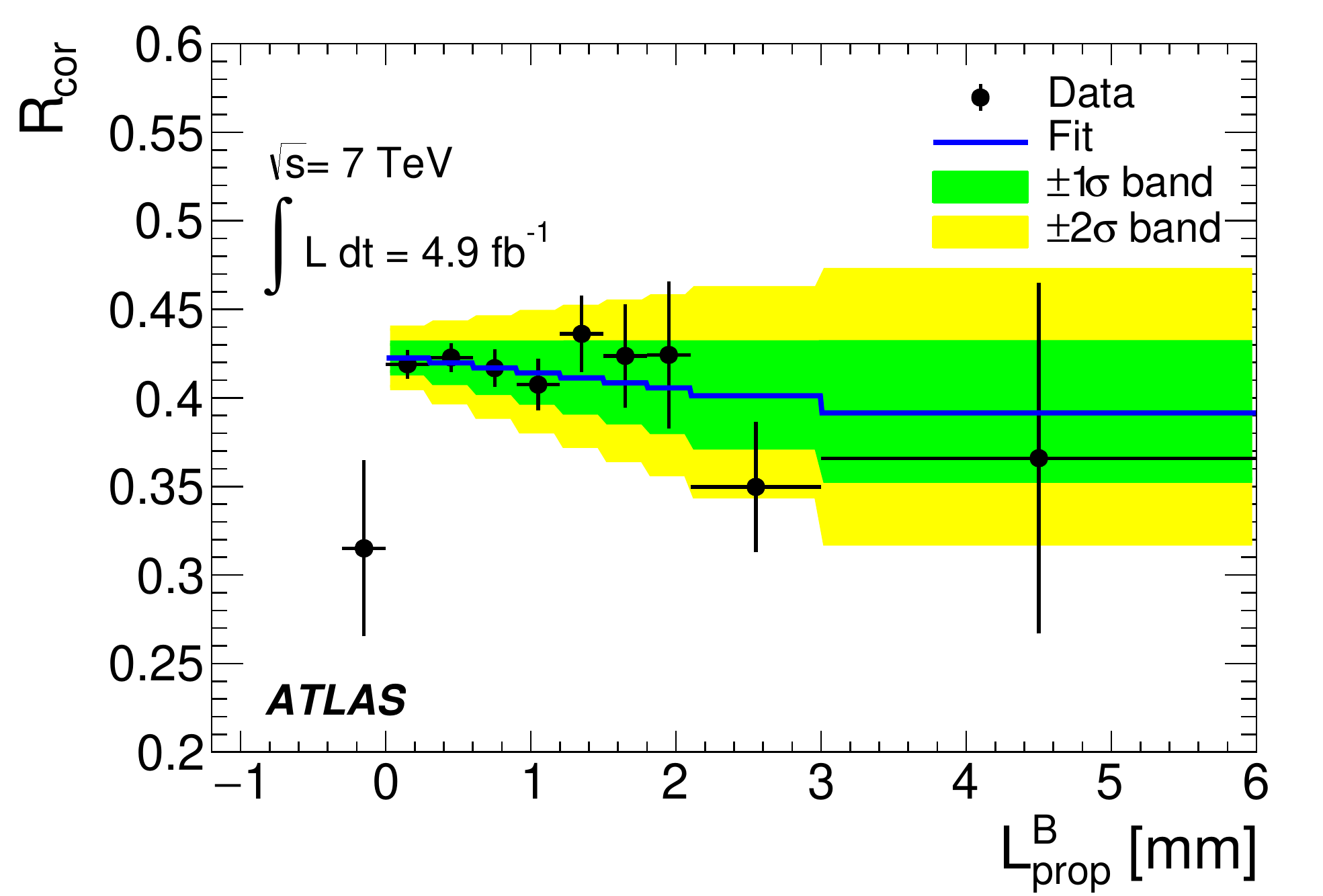}
    }
    \subfigure[]{
      \hspace*{-0.5cm}    
      \includegraphics[width=0.52\textwidth]{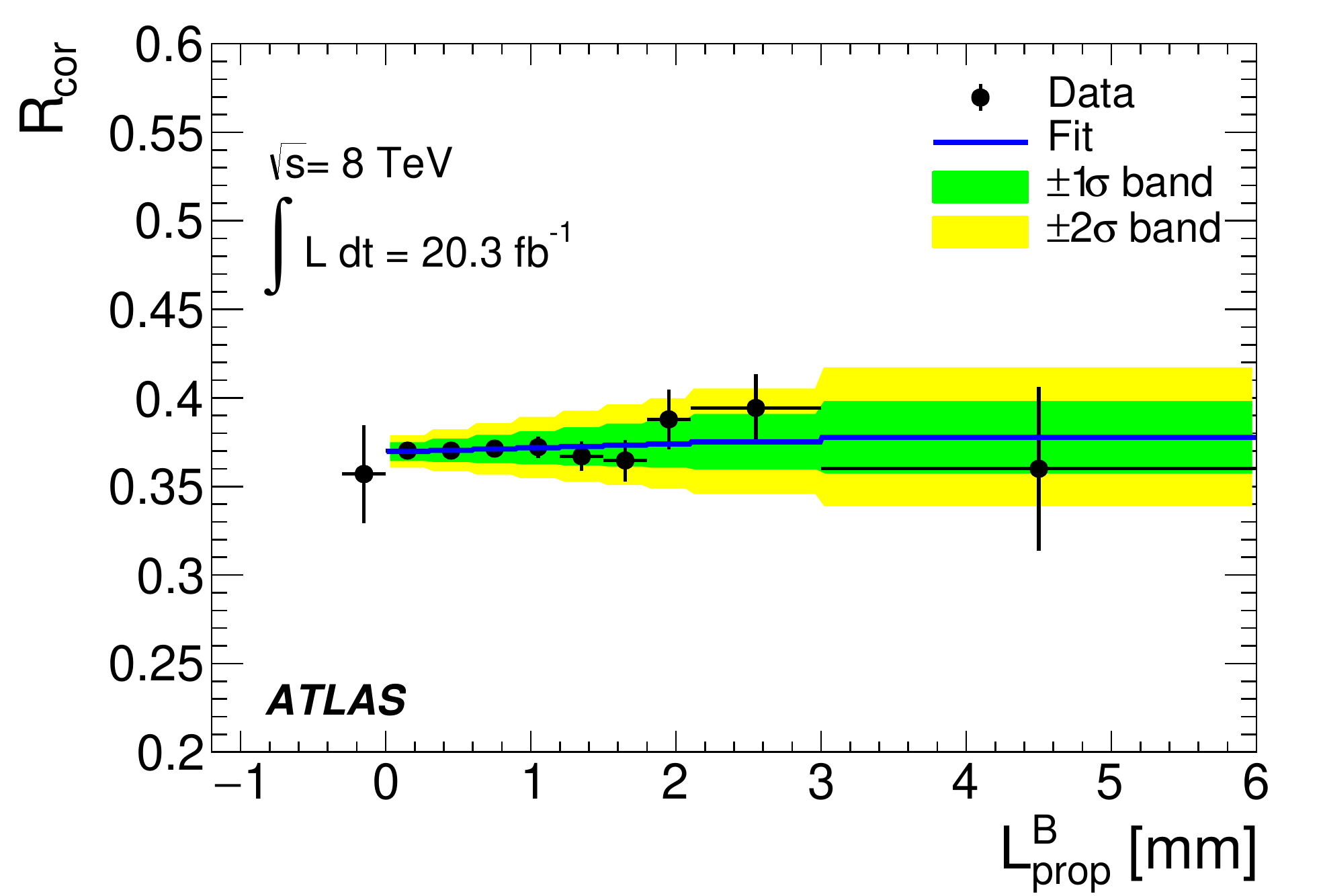}
      \hspace*{-0.6cm}
    }
    \caption{Efficiency-corrected ratio of the observed decay length distributions,
    $R_{\mathrm{cor}}(\lprop)$ for (a) $\sqrt{s} = 7$ TeV 
    and (b) $\sqrt{s} = 8$ TeV data sets.
    The normalisation of the two data sets is arbitrary. The full line
    shows the fit of $R_{\mathrm{cor}}(\lprop)$ to $R_{\mathrm{exp}}$ given by Eq. (\ref{rexp}).
    The error bands correspond to uncertainties in $\Delta \Gamma_d / \Gamma_d$ determined by the fit.}
    \label{fig_rexp}
  \end{center}
\end{figure}

The obtained values of $R_{i,{\mathrm{eff}}}$ are used to correct
the observed ratio $R_{i,{\mathrm{uncor}}}$ given by Eq. (\ref{rlp}).
The resulting ratio $R_{i,{\mathrm{cor}}}$ is defined as:
\begin{equation}
\label{r_cor}
R_{i,{\mathrm{cor}}} = \frac{R_{i,{\mathrm{uncor}}}}{R_{i,{\mathrm{eff}}}}.
\end{equation}
This ratio is shown in Figure \ref{fig_rexp}. It is used to obtain $\Delta \Gamma_d /\Gamma_d$ by the following procedure.
For each $\lprop$ bin $i$ defined in Table \ref{tab2}, the expected numbers of events in the $\jpsi K_S$
and $\jpsi K^{*0}$ channels are computed as:
\begin{eqnarray}
N_i[\Delta \Gamma_d/\Gamma_d,\jpsi K_S] & = & C_1 \int_{L^{\mathrm{min}}_i}^{L^{\mathrm{max}}_i} \Gamma[\lprop, \jpsi K_S] \mathrm{d} \lprop, \\
N_i[\Delta \Gamma_d/\Gamma_d, \jpsi K^{*0}] & = & C_2 \int_{L^{\mathrm{min}}_i}^{L^{\mathrm{max}}_i} \Gamma[\lprop, \jpsi K^{*0}] \mathrm{d} \lprop.
\end{eqnarray}
The integration limits $L^{\mathrm{min}}_i$ and $L^{\mathrm{max}}_i$ for each bin $i$ are given by
the lower and upper bin edges in Table \ref{tab2}. $C_1$ and $C_2$
are arbitrary normalisation coefficients.
The expressions for $\Gamma[\lprop, J/\psi K_S]$ and $\Gamma[\lprop, J/\psi K^{*0}]$ are given by
Eqs. (\ref{glcp}) and (\ref{glfs}), respectively. The sensitivity to $\Delta \Gamma_d$ comes
from $\Gamma[\lprop, J/\psi K_S]$ (see Eq. (\ref{gcp}))
while $\Gamma[\lprop, J/\psi K^{*0}]$ provides the normalisation,
which helps to reduce the systematic uncertainties.

The expected ratio of the decay rates in the two channels in each $\lprop$ bin is:
\begin{equation}
\label{rexp}
R_{i, {\mathrm{exp}}}[\Delta \Gamma_d/\Gamma_d] = \frac{N_i[\Delta \Gamma_d/\Gamma_d, \jpsi K_S]}{N_i[\Delta \Gamma_d/\Gamma_d, \jpsi K^{*0}]}.
\end{equation}
The relative width difference $\Delta \Gamma_d / \Gamma_d$ is obtained
from a $\chi^2$ minimisation:
\begin{equation}
\chi^2[\Delta \Gamma_d/\Gamma_d] = \sum_{i=2}^{10} \frac{(R_{i,{\mathrm{cor}}} - R_{i,{\mathrm{exp}}}[\Delta \Gamma_d/\Gamma_d])^2}{\sigma_i^2}.
\end{equation}
The values $\sigma_i$ are the statistical
uncertainties of $R_{i,{\mathrm{cor}}}$.
In the sum, the first
bin of $\lprop$ is not included as it corresponds to a negative decay length.

The free parameters in this minimisation are the overall normalisation and $\Delta \Gamma_d /\Gamma_d$. All other
parameters describing the $\Bd$ meson are fixed to their world average values.
The fit is performed
separately for the 2011 and 2012 samples because the systematic uncertainties for the two data samples are different.
The result of the fit is shown in Figure \ref{fig_rexp}.
The $\chi^2$ of the fit is 4.34 (NDF = 7) in the 2011 data set and 2.81 (NDF = 7) 
in the 2012 data set.

The fit yields
\begin{eqnarray}
\label{yd_2011}
\Delta \Gamma_d / \Gamma_d  & = & (-2.8 \pm 2.2~\mbox{(stat.)} \pm 1.5~\mbox{(MC stat.)}) \times 10^{-2}~~~~\mbox{(2011)}, \\
\label{yd_2012}
\Delta \Gamma_d / \Gamma_d  & = & (+0.8 \pm 1.3~\mbox{(stat.)} \pm 0.5~\mbox{(MC stat.)}) \times 10^{-2}~~~~\mbox{(2012)}.
\end{eqnarray}
Here the uncertainties due to the data and MC statistics are given separately.
The systematic uncertainties are discussed in Section \ref{sec_syst}.

\FloatBarrier

\section{Systematic uncertainties}
\label{sec_syst}

The relative $\Bd$ width difference is extracted from the ratio of the $\lprop$ distributions in the two $\Bd$ decay modes, which
are obtained using a similar procedure, the same type of information and in the same production environment.
Therefore, the impact of many systematic uncertainties, such as the trigger selection, decay-time resolution
or $\Bd$ production properties, is negligible. However, some differences between the $\Bd \to \jpsi K_S$ and
$\Bd \to \jpsi K^{*0}$ channels cannot be eliminated and the inaccuracy of their simulation results in
systematic uncertainties, which are estimated in this section.

The mean proper decay length of the $K_S$ meson is $26.8$ mm.
Since the $p_\mathrm{T}$ of the $K_S$ meson can be high, some $K_{s}$ mesons decay outside the inner detector and are lost.
The probability of losing a $K_S$ meson is higher for large $\Bd$ decay time due to the reduction of the fiducial
volume of the $K_S$ decay.
Thus, the displaced vertex of the $K_S$ decay and the absence of such a vertex in the $K^{*0} \to K^+ \pi^-$
decay results in a decay-time dependence of $R_{\mathrm{eff}}$ defined in Eq. (\ref{r_eff}). 
Applying the correction given by Eq. (\ref{r_cor}) to $R_{\mathrm{uncor}}$ 
takes into account this dependence.

The test of the simulated $K_S$ reconstruction is performed by comparing the distribution of the $K_S$ decay length
and the $K_S$ pseudorapidity in data and simulation.
This dedicated study shows that there is a residual difference between data and MC simulation in the distributions of the
laboratory decay length of reconstructed $K_S$ mesons projected along the $K_S$ momentum in the transverse plane,
$L_{xy}(K_S)$. It is caused by the remaining difference between
the $K_S$ momentum distributions in data and MC simulation.
After applying an additional weight to the Monte Carlo events to correct for this
difference a change in the value of $\Delta \Gamma_d /\Gamma_d$ of $\delta(\Delta \Gamma_d /\Gamma_d) = -0.21 \times 10^{-2}$
is obtained for the 2011 data set and $\delta(\Delta \Gamma_d /\Gamma_d) = -0.16 \times 10^{-2}$
for the 2012 data set. This difference is taken as the systematic uncertainty due to modelling of the
$L_{xy}(K_S)$ dependence of the $K_S$ reconstruction.

The same procedure is applied for the pseudorapidity distribution of the $K_S$ meson, $\eta(K_S)$.
The systematic uncertainty due to modelling of the $|\eta(K_S)|$ dependence
of the $K_S$ reconstruction is estimated by re-weighting the MC events to make the $|\eta(K_S)|$ distribution
the same as in data.
The observed changes are  $\delta(\Delta \Gamma_d /\Gamma_d) = +0.14 \times 10^{-2}$ for the 2011 data set
and $\delta(\Delta \Gamma_d /\Gamma_d) = -0.01 \times 10^{-2}$ for the 2012 data set.

The systematic uncertainty due to the choices made in the model used to fit the mass distributions can be estimated by
considering different variations of the fit model. The range over which
the $B^{0} \rightarrow \jpsi K_{S}$ and $B^{0} \rightarrow \jpsi K^{*0}$ mass fits are applied is varied
and the measurement of $\Delta \Gamma_d /\Gamma_d$
is repeated for each variation. The systematic uncertainty is estimated by taking
the difference between the values of $\Delta \Gamma_d /\Gamma_d$
obtained from the default fit and each of the varied fits.
Variations $\delta(\Delta \Gamma_d /\Gamma_d) = -0.47 \times 10^{-2}$ and $-0.30 \times 10^{-2}$
are obtained for the 2011 data set in the $\jpsi K_S$ and $\jpsi K^{*0}$ channels, respectively.
The changes for the 2012 data set are  $\delta(\Delta \Gamma_d /\Gamma_d) = -0.59 \times 10^{-2}$
and $-0.15 \times 10^{-2}$ in the $\jpsi K_S$ and $\jpsi K^{*0}$ channels, respectively.
These values are included as the systematic uncertainty from this source.

Additionally, the background function is changed from an exponential to a fourth-order polynomial and the
systematic uncertainty due to the choice of background function is estimated from the difference between
the value of $\Delta \Gamma_d /\Gamma_d$
from the default fit and the value from the fit using the polynomial background function.
A change $\delta(\Delta \Gamma_d /\Gamma_d) = -0.16 \times 10^{-2}$ is obtained for the 2011 data set.
The change for the 2012 data set is  $\delta(\Delta \Gamma_d /\Gamma_d) = +0.09 \times 10^{-2}$.

In the fit of the number of $\Bd \to \jpsi K_S$ decays the contribution from the $\Bs \to \jpsi K_S$ is
a free parameter of the fit. As a systematic uncertainty cross-check, the ratio of the yields of these
two decays is fixed to be the same as that measured by the LHCb Collaboration \cite{bslt-lhcb}. The resulting
change in the $\Delta \Gamma_d /\Gamma_d$ value is $\delta(\Delta \Gamma_d /\Gamma_d) = -0.11 \times 10^{-2}$
for the 2011 data set and $\delta(\Delta \Gamma_d /\Gamma_d) = +0.08 \times 10^{-2}$ for the 2012 data set
and is included as an additional source of systematic uncertainty.

The systematic uncertainty due to the resolution of $\lprop$ is also considered.
The average decay length resolution is 35 $\mu$m for $\Bs \to \jpsi K_S$ and 33 $\mu$m for $\Bd \to \jpsi K^{*0}$.
In this analysis, separate resolution functions are used for the 
two channels. To test the sensitivity to the resolution, the
measurement of $\Delta \Gamma_d /\Gamma_d$ is repeated by using the resolution of $\jpsi K^{*0}$
for both channels.
A change in the value of $\Delta \Gamma_d /\Gamma_d$ of $\delta(\Delta \Gamma_d /\Gamma_d) = -0.29 \times 10^{-2}$
is obtained and is used as the systematic uncertainty from this source. It is found to be the same
for the 2011 and 2012 data sets.

A toy MC sample is employed to identify any possible bias in the
fitting procedure.
In this toy MC sample, the expected number of $\jpsi K_S$ and $\jpsi K^{*0}$ candidates
in each bin of $\lprop$ is
determined according to the analytic functions given by Eqs. (\ref{gcp}) and (\ref{gfs}), respectively, and
a value of $\Delta\Gamma_d/\Gamma_d = 0.42 \times 10^{-2}$ corresponding to the SM expectation
\cite{Lenz}.
Using these expected numbers of candidates as the mean values,
the number of candidates in both channels is randomly generated in each $\lprop$
bin with an uncertainty corresponding to that obtained in data.
The ratio of the obtained distributions is then fitted using the method described in Section \ref{sec_rcorr}.
The procedure is repeated $10\,000$ times 
giving a bias in the mean fitted value 
$\delta(\Delta \Gamma_d /\Gamma_d) = +0.07 \times 10^{-2}$.
This value is used as the systematic uncertainty due to the fitting procedure and
it is taken to be the same for the 2011 and 2012 data sets.

The impact of the uncertainty of the $\Bd$ production asymmetry
is
$\delta(\Delta \Gamma_d /\Gamma_d) = 0.01 \times 10^{-2}$
for both the 2011 and 2012 data sets.

The systematic uncertainty from the number of events in the MC samples
corresponds to an uncertainty of $\delta(\Delta \Gamma_d /\Gamma_d) = 1.54 \times 10^{-2}$
for the 2011 data set and $\delta(\Delta \Gamma_d /\Gamma_d) = 0.45 \times 10^{-2}$ for the 2012 data set.

Table \ref{tab_syst} gives a summary of the estimated systematic uncertainties.
All of the quantified systematic uncertainties are symmetrized.

In addition to the estimate of the systematic uncertainty, several cross-checks
are performed. Some of the selection cuts described in Section 6 are modified and
the corresponding changes in the $\Delta \Gamma_d /\Gamma_d$ value are assessed.
In particular,
the transverse momenta of the charged pions from the $K_{S}$ decay and the charged pion
from the $K^{*0}$ decay are required to be greater than 500 MeV, rather than 400 MeV.
%
Also, the transverse momentum of the charged kaon from the $K^{*0}$ is required to be greater than
1 GeV, rather than 800 MeV.
%
Additionally, the transverse momentum of the $\Bd$ meson is required to be less than 60 GeV.
In all cases, the change of the measured value of $\Delta \Gamma_d$ is consistent with
fluctuations due to the reduced number of events.

Furthermore, a number of consistency checks related to the description of the experimental conditions in simulation
are performed. Most notably, the MC description of the spread of the $z$ position of the primary vertex,
the angular distributions of the $\Bd$ decay products, and the trigger rates are studied in detail.
In all cases, the residual differences between data and MC simulation do not impact the measured value of $\Delta \Gamma_d$.


\begin{table}[htb]
  \begin{center}
    \begin{tabular}{c c c}
      \hline \hline
      Source & $\delta(\Delta \Gamma_d /\Gamma_d)$, 2011 & $\delta(\Delta \Gamma_d /\Gamma_d)$, 2012  \\
      \hline
      $K_S$ decay length                            & $ 0.21 \times 10^{-2}$ & $ 0.16 \times 10^{-2}$ \\
      $K_S$ pseudorapidity                          & $ 0.14 \times 10^{-2}$ & $ 0.01 \times 10^{-2}$ \\
      $\Bd \rightarrow \jpsi K_{S}$ mass range      & $ 0.47 \times 10^{-2}$ & $ 0.59 \times 10^{-2}$ \\
      $\Bd \rightarrow \jpsi K^{*0}$ mass range     & $ 0.30 \times 10^{-2}$ & $ 0.15 \times 10^{-2}$ \\
      Background description                        & $ 0.16 \times 10^{-2}$ & $ 0.09 \times 10^{-2}$ \\
      $\Bs \to \jpsi K_S$ contribution              & $ 0.11 \times 10^{-2}$ & $ 0.08 \times 10^{-2}$ \\
      $L_{prop}^{B}$ resolution                     & $ 0.29 \times 10^{-2}$ & $ 0.29 \times 10^{-2}$ \\
      Fit bias (Toy MC)                             & $ 0.07 \times 10^{-2}$ & $ 0.07 \times 10^{-2}$ \\
      $\Bd$ production asymmetry                    & $ 0.01 \times 10^{-2}$ & $ 0.01 \times 10^{-2}$ \\
      MC sample                                     & $ 1.54 \times 10^{-2}$ & $ 0.45 \times 10^{-2}$ \\
      \hline
      Total uncertainty                             & $ 1.69 \times 10^{-2}$ & $ 0.84 \times 10^{-2}$ \\
      \hline \hline
    \end{tabular}
    \caption{Sources of systematic uncertainty in the $\Delta \Gamma_d /\Gamma_d$ measurement and their values for the 2011 and 2012 data sets.}
    \label{tab_syst}
  \end{center}
\end{table}

\FloatBarrier

\section{Results}
\label{sec_result}

Using the measurements of $\Delta \Gamma_d / \Gamma_d$ given in Eqs. (\ref{yd_2011}) and (\ref{yd_2012})
and the study of systematic uncertainties presented in Section \ref{sec_syst}, the following measurements are obtained:
\begin{eqnarray}
\label{yd_2011_all}
\Delta \Gamma_d / \Gamma_d  & = & (-2.8 \pm 2.2~\mbox{(stat.)} \pm 1.7~\mbox{(syst.)}) \times 10^{-2}~~~~\mbox{(2011)}, \nonumber \\
\label{yd_2012_all}
\Delta \Gamma_d / \Gamma_d  & = & (+0.8 \pm 1.3~\mbox{(stat.)} \pm 0.8~\mbox{(syst.)}) \times 10^{-2}~~~~\mbox{(2012)} \nonumber.
\end{eqnarray}
In the combination of these measurements, the correlations of different sources
of systematic uncertainty between the two years are taken into account.
The systematic uncertainties due to the background description
and the size of the MC samples
are assumed to be uncorrelated. All other sources of systematic uncertainty
are taken to be fully correlated.
The combination is done using the $\chi^2$ method.
The $\chi^2$ function includes the correlation terms of the different components of the uncertainty as specified above.
The combined result for the data collected by the ATLAS experiment in Run 1 is:
\begin{equation}
\Delta \Gamma_d / \Gamma_d  = (-0.1 \pm 1.1~\mbox{(stat.)} \pm 0.9~\mbox{(syst.)}) \times 10^{-2} \nonumber.
\end{equation}
It is currently the most precise single measurement
of this quantity. It agrees well with the
SM prediction~\cite{Lenz}
and is consistent with other measurements of this quantity \cite{DG-Belle,DG-LHCB,dg-babar-2}.
It also agrees with the indirect measurement by the D0 Collaboration \cite{dg-d0}.

\section{Conclusions}
  The measurement of the relative width difference $\Delta \Gamma_d / \Gamma_d$
  of the $B^0$--$\bar B^0$ system is performed using the data collected by the ATLAS experiment at the LHC in $p p$
  collisions at $\sqrt{s} = 7$ TeV and $\sqrt{s} = 8$ TeV and
  corresponding to an integrated luminosity of 25.2 fb$^{-1}$.
  The value of $\Delta \Gamma_d / \Gamma_d$ is obtained by comparing the decay time distributions
  of $B^0 \to J/\psi K_S$ and $B^0 \to J/\psi K^{*0}(892)$ decays. The result is
  \begin{equation}
\Delta \Gamma_d / \Gamma_d  = (-0.1 \pm 1.1~\mbox{(stat.)} \pm 0.9~\mbox{(syst.)}) \times 10^{-2} \nonumber.
  \end{equation}
  Currently, this is the most precise single
  measurement of $\Delta \Gamma_d / \Gamma_d$. It agrees with the Standard Model
  prediction and the measurements by other experiments.

The production
asymmetry of the $\Bd$ meson with $p_\mathrm{T}(\Bd) > 10$ GeV and $|\eta(\Bd)| < 2.5$ is found to be
\begin{equation}
A_{\mathrm P}(\Bd) = (+0.25 \pm 0.48 \pm 0.05) \times 10^{-2} \nonumber .
\end{equation}
The value of $A_{\mathrm P}(\Bd)$ is consistent with the measurement of the LHCb Collaboration
performed in the  $2.5 < \eta(\Bd) < 4.0$ and $ 4 < p_{\mathrm{T}}(\Bd) < 30$~GeV range.

\section*{Acknowledgements}


We thank CERN for the very successful operation of the LHC, as well as the
support staff from our institutions without whom ATLAS could not be
operated efficiently.

We acknowledge the support of ANPCyT, Argentina; YerPhI, Armenia; ARC, Australia; BMWFW and FWF, Austria; ANAS, Azerbaijan; SSTC, Belarus; CNPq and FAPESP, Brazil; NSERC, NRC and CFI, Canada; CERN; CONICYT, Chile; CAS, MOST and NSFC, China; COLCIENCIAS, Colombia; MSMT CR, MPO CR and VSC CR, Czech Republic; DNRF and DNSRC, Denmark; IN2P3-CNRS, CEA-DSM/IRFU, France; GNSF, Georgia; BMBF, HGF, and MPG, Germany; GSRT, Greece; RGC, Hong Kong SAR, China; ISF, I-CORE and Benoziyo Center, Israel; INFN, Italy; MEXT and JSPS, Japan; CNRST, Morocco; FOM and NWO, Netherlands; RCN, Norway; MNiSW and NCN, Poland; FCT, Portugal; MNE/IFA, Romania; MES of Russia and NRC KI, Russian Federation; JINR; MESTD, Serbia; MSSR, Slovakia; ARRS and MIZ\v{S}, Slovenia; DST/NRF, South Africa; MINECO, Spain; SRC and Wallenberg Foundation, Sweden; SERI, SNSF and Cantons of Bern and Geneva, Switzerland; MOST, Taiwan; TAEK, Turkey; STFC, United Kingdom; DOE and NSF, United States of America. In addition, individual groups and members have received support from BCKDF, the Canada Council, CANARIE, CRC, Compute Canada, FQRNT, and the Ontario Innovation Trust, Canada; EPLANET, ERC, FP7, Horizon 2020 and Marie Sk{\l}odowska-Curie Actions, European Union; Investissements d'Avenir Labex and Idex, ANR, R{\'e}gion Auvergne and Fondation Partager le Savoir, France; DFG and AvH Foundation, Germany; Herakleitos, Thales and Aristeia programmes co-financed by EU-ESF and the Greek NSRF; BSF, GIF and Minerva, Israel; BRF, Norway; Generalitat de Catalunya, Generalitat Valenciana, Spain; the Royal Society and Leverhulme Trust, United Kingdom.

The crucial computing support from all WLCG partners is acknowledged
gratefully, in particular from CERN and the ATLAS Tier-1 facilities at
TRIUMF (Canada), NDGF (Denmark, Norway, Sweden), CC-IN2P3 (France),
KIT/GridKA (Germany), INFN-CNAF (Italy), NL-T1 (Netherlands), PIC (Spain),
ASGC (Taiwan), RAL (UK) and BNL (USA) and in the Tier-2 facilities
worldwide.

\printbibliography
\clearpage

\newpage 
\begin{flushleft}
{\Large The ATLAS Collaboration}

\bigskip

M.~Aaboud$^\textrm{\scriptsize 136d}$,
G.~Aad$^\textrm{\scriptsize 87}$,
B.~Abbott$^\textrm{\scriptsize 114}$,
J.~Abdallah$^\textrm{\scriptsize 65}$,
O.~Abdinov$^\textrm{\scriptsize 12}$,
B.~Abeloos$^\textrm{\scriptsize 118}$,
R.~Aben$^\textrm{\scriptsize 108}$,
O.S.~AbouZeid$^\textrm{\scriptsize 138}$,
N.L.~Abraham$^\textrm{\scriptsize 150}$,
H.~Abramowicz$^\textrm{\scriptsize 154}$
% ...
\bigskip
\\
% ...
$^{12}$ Institute of Physics, Azerbaijan Academy of Sciences, Baku, Azerbaijan\\
% ...
$^{65}$ University of Iowa, Iowa City IA, United States of America\\
% ...
$^{87}$ CPPM, Aix-Marseille Universit{\'e} and CNRS/IN2P3, Marseille, France\\
% ...
$^{108}$ Nikhef National Institute for Subatomic Physics and University of Amsterdam, Amsterdam, Netherlands\\
% ...
$^{114}$ Homer L. Dodge Department of Physics and Astronomy, University of Oklahoma, Norman OK, United States of America\\
% ...
$^{118}$ LAL, Univ. Paris-Sud, CNRS/IN2P3, Universit{\'e} Paris-Saclay, Orsay, France\\
% ...
$^{136}$ $^{(a)}$ Facult{\'e} des Sciences Ain Chock, R{\'e}seau Universitaire de Physique des Hautes Energies - Universit{\'e} Hassan II, Casablanca; $^{(b)}$ Centre National de l'Energie des Sciences Techniques Nucleaires, Rabat; $^{(c)}$ Facult{\'e} des Sciences Semlalia, Universit{\'e} Cadi Ayyad, LPHEA-Marrakech; $^{(d)}$ Facult{\'e} des Sciences, Universit{\'e} Mohamed Premier and LPTPM, Oujda; $^{(e)}$ Facult{\'e} des sciences, Universit{\'e} Mohammed V, Rabat, Morocco\\
% ...
$^{138}$ Santa Cruz Institute for Particle Physics, University of California Santa Cruz, Santa Cruz CA, United States of America\\
% ...
$^{150}$ Department of Physics and Astronomy, University of Sussex, Brighton, United Kingdom\\
% ...
$^{154}$ Raymond and Beverly Sackler School of Physics and Astronomy, Tel Aviv University, Tel Aviv, Israel\\
% ...
$^{a}$ Also at Department of Physics, King's College London, London, United Kingdom\\
$^{b}$ Also at Institute of Physics, Azerbaijan Academy of Sciences, Baku, Azerbaijan\\
$^{c}$ Also at Novosibirsk State University, Novosibirsk, Russia\\
$^{d}$ Also at TRIUMF, Vancouver BC, Canada\\
$^{e}$ Also at Department of Physics {\&} Astronomy, University of Louisville, Louisville, KY, United States of America\\
% ...
\end{flushleft}


\end{document}